\documentclass[aps,prl,twocolumn,showpacs,preprintnumbers,amsmath,amssymb]{revtex4}
\usepackage{graphicx}
\usepackage{graphicx}
\usepackage{dcolumn}
\usepackage{bm}
\newcommand{\be}{\begin{eqnarray}}
\newcommand{\ee}{\end{eqnarray}}

 \newcommand{\gsim}{\mathrel{\hbox{\rlap{\lower.55ex \hbox {$\sim$}}
                   \kern-.3em \raise.4ex \hbox{$>$}}}}
\newcommand{\lsim}{\mathrel{\hbox{\rlap{\lower.55ex \hbox {$\sim$}}
                   \kern-.3em \raise.4ex \hbox{$<$}}}}

\newcommand{\ave}[1]{\left\langle #1 \right\rangle}

\def\roughly#1{\mathrel{\raise.3ex\hbox{$#1$\kern-.75em%
\lower1ex\hbox{$\sim$}}}}
\def\lsim{\roughly<}
\def\gsim{\roughly>}

\begin{document}

\title{Exposing the non-collectivity in elliptic flow }

\author{Jinfeng Liao}
\email{jliao@lbl.gov}
\author{Volker Koch}
\email{vkoch@lbl.gov}
\affiliation{Nuclear Science Division, Lawrence Berkeley National Laboratory, MS70R0319, 1 Cyclotron Road, Berkeley, CA 94720}




\begin{abstract}
We show that backward-forward elliptic anisotropy correlation
provides an experimentally accessible observable which
distinguishes between collective and non-collective contributions
to the observed elliptic anisotropy $v_2$ in relativistic heavy
ion collisions. The measurement of this observable will reveal the
momentum scale at which collective expansion seizes and where the
elliptic anisotropy is dominated by (semi)-hard processes.

\end{abstract}


\maketitle

Deconfined QCD matter at high energy density, the so-called
Quark-Gluon Plasma (QGP), was a phase during the evolution of the
early universe and it is now created and explored experimentally
in relativistic heavy ion collisions. One major discovery by the
experimental program \cite{rhic_white_paper} of the Relativistic
Heavy Ion Collider (RHIC) is the large elliptic
anisotropy\cite{Voloshin:2008dg}, $v_2$, of observed particles'
transverse momenta, which is consistent with predictions from
ideal hydrodynamic expansion \cite{Teaney:2000cw}. This led to the
conjecture that the matter created in these collisions exhibits
``perfect fluidity'', i.e. minimal shear viscosity. First studies
\cite{Csernai:2006zz,Lacey:2006bc,viscous_hydro} indicate that the
shear viscosity  must be very small especially near $T_c$, and
likely smaller than any known condensed matter substances and
rather close to the conjectured universal lower bound
$\frac{\eta}{s}\geq\frac{1}{4\pi}$ based on calculations utilizing
the gauge/string duality \cite{lower_bound}.

\begin{figure*}
  \hskip 0in\includegraphics[width=4.7cm]{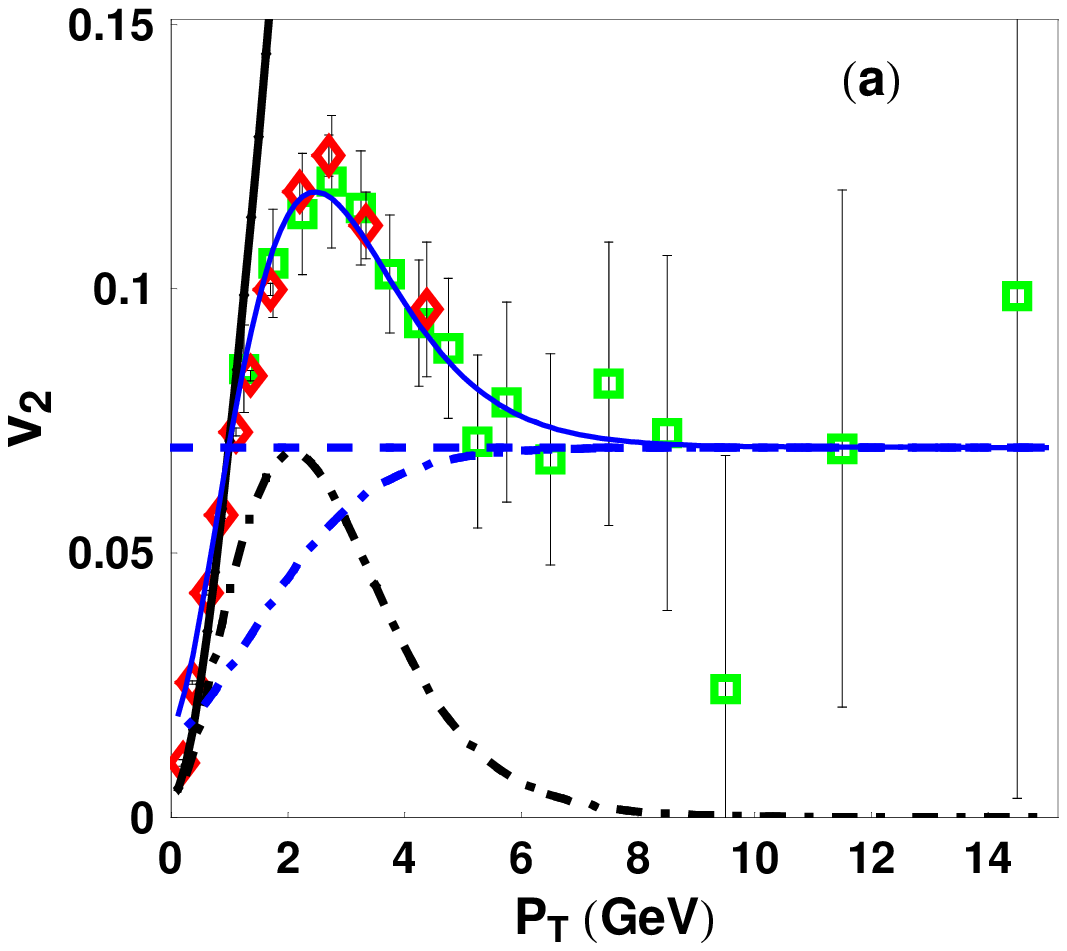}
  \hskip 0.1in\includegraphics[width=4.7cm]{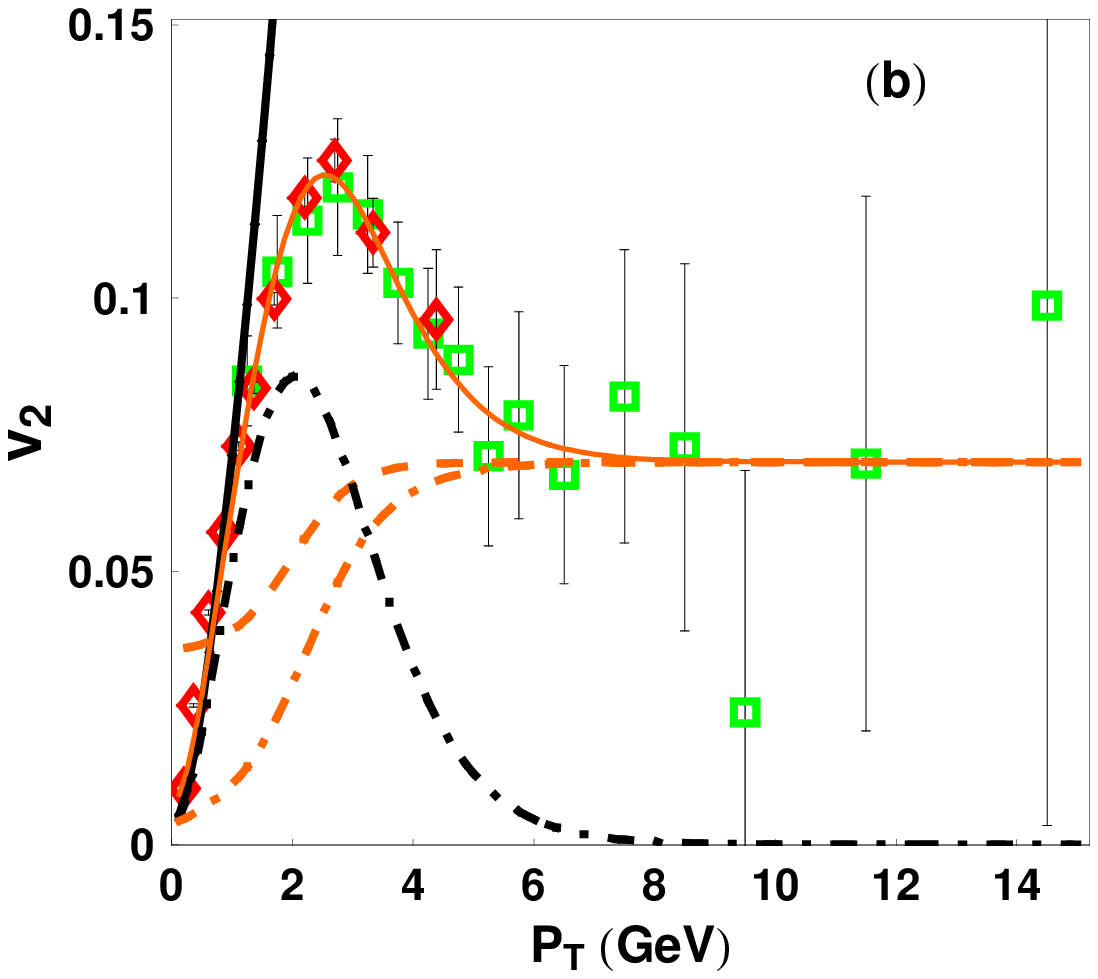}
  \hskip 0.1in\includegraphics[width=4.7cm]{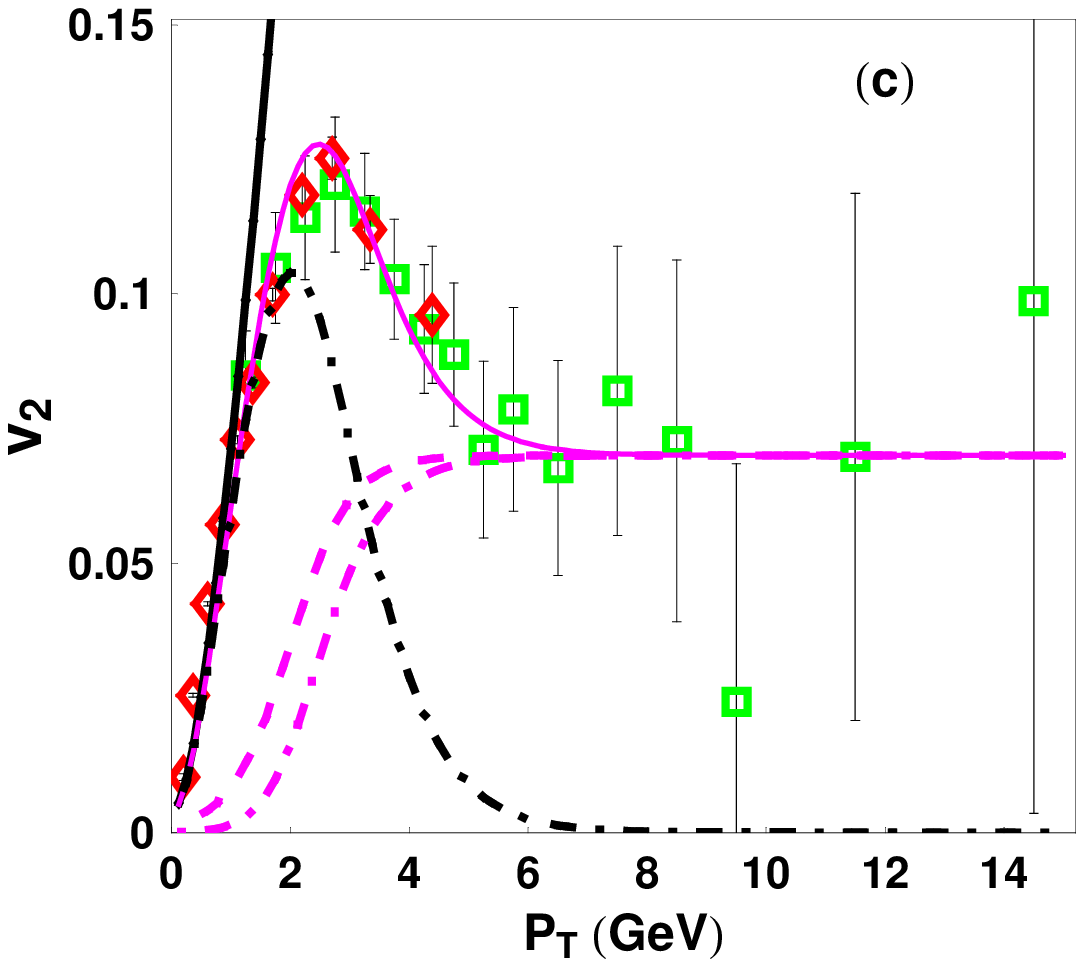}
 \caption{\label{fig_v2} Parametrization of $v_2$ data for AuAu $200\, \rm GeV$, (00-20\% centrality). The red diamonds are for negative charged hadrons  \protect\cite{Adler:2003kt} and
 the green boxes are for $\pi_0$ from PHENIX Run7 Preliminary
 \protect\cite{Adler:2006bw,phenix_prel}. The blue/orange/magenta
 solid lines in (a)/(b)/(c) represent the combined
 parametrization, Eq.(\ref{eqn_v2_combo}), respectively. In all three panels, the black solid lines
 show the collective-flow-only contribution while the colored long-dashed lines (extending to high $p_t$) show the jet-only
 contribution in Eq.(\ref{eqn_hpt_vt}).
 The dash-dotted lines show the weighted contributions $(1-g)\cdot <v_2^{\mathcal F}>$ (black)
 and $\, g\cdot <v_2^{\mathcal J}>$ (colored), respectively. }
 \end{figure*}

The transverse momentum ($p_t$) dependence of the elliptic
anisotropy, $v_2$, as depicted in Fig. \ref{fig_v2} shows a rise at
low $p_t$ towards a maximum at $p_t\simeq 3\,\rm GeV$, and a
constant value for large transverse momenta. At present the rise
at low transverse momentum ($p_t\lesssim 1.5\,\rm GeV$) is thought
to be due to \emph{collective} hydrodynamic expansion, which
translates the initial spatial anisotropy into a $p_t$ anisotropy
\cite{Ollitrault:1992bk} over a wide region in pseudo-rapidity.

At very high $p_t$, on the other hand, $v_2$ is believed to result
from the different attenuation of hard partons (jet quenching) in
the asymmetrically distributed matter
\cite{Shuryak:2001me,Gyulassy:2000gk,Liao:2008dk}. In this case
the elliptic anisotropy, $v_2$, is \emph{non-collective} in the
sense that it is due to jet-like processes which are rather local
in rapidity \cite{Noncollectivity}. The transition from collective
(hydrodynamic) to non-collective (jet-like) anisotropy is expected
to take place around $p_t \simeq 2-4 \,\rm Gev$, but the details
are not well understood. In addition, one would expect that the
attenuation of the jets should result in local, non-collective
anisotropy also at lower $p_t$, since the debris from the
jet-quenching process needs to go somewhere. If there is indeed a
sizeable non-collective contribution to $v_2$ at low $p_t$,
comparisons of (viscous) hydrodynamics with the data may lead to
wrong conclusions about the viscosity of the produced matter in
these collisions. Therefore, it would be desirable to have a
direct measurement not only of the $p_t$ scale at which the
transition from collective to non-collective $v_2$ occurs, but
also of the contribution of non-collective effects in the low
$p_t$ region, $p_t \lesssim 1.5 \,\rm GeV$.

It is the purpose of this work to propose an observable which will
distinguish between the collective and the non-collective
contributions to the elliptic anisotropy. In this paper, for
illustration purposes, we will identify the collective component
with hydrodynamic flow and the non-collective one with attenuated
jets. We note that there may be other mechanisms at work which
generate similar collective and/or non-collective contributions to
the elliptic anisotropy, and the proposed observable is not
sensitive to any specific mechanism.

The key observation is the following: consider two rapidity bins,
one forward, one backward, with a suitable separation (gap) in
between. In a given event, the elliptic anisotropy of the
collective component in both rapidity bins is highly correlated.
The non-collective component, on the other hand, generates an
elliptic anisotropy in either the forward bin or the backward bin,
but never in both, resulting in very small backward-forward
correlations. To illustrate this point, consider hydrodynamic
expansion as an example for the collective component. In each
event the elliptic anisotropy is aligned in azimuth and of
comparable magnitude in the forward and backward bins. Contrast
this with the elliptic anisotropy due to quenched jets, as an
example for a non-collective component. At RHIC energies one
rarely has more than one hard process per event. In addition, the
anisotropy due to attenuated jets will be very local in rapidity
and, therefore, will either contribute to the forward or to the
backward bin but not to both. Consequently, these processes do not
give rise to long-range forward-backward correlation of the
elliptic anisotropy. At higher energies, such as the Large Hadron
Collider (LHC) with many hard processes per event one has to
choose a suitable rapidity gap, which is larger than typical jet
cone extension but smaller than average jet-jet separation. After
these general considerations, we will proceed to define the
proposed observable and demonstrate its sensitivity to the degree
of collectivity in contributions to $v_2$.

Let us start by recalling the definition of $v_2$:
\begin{equation} \label{eqn_vt_def}
\ave{v_2(p_t)}= \frac{ \int_0^{2\pi} d\phi \cos(2\phi) \ave{
\frac{d^2 N}{p_t d p_t d\phi } }}{\int_0^{2\pi} d\phi \ave{
\frac{d^2 N}{p_t d p_t d\phi} }} \equiv \frac{\ave{
V_2(p_t)}}{\ave{\frac{d N}{p_t d p_t} }}
\end{equation}
In the above $\ave{\frac{d^2 N}{p_t d p_t d\phi}}$ is the
distribution of the event-averaged $p_t$-differential yield over
the azimuthal angle $\phi$ which is defined with respect to the
reaction plane.  The numerator, denoted as $V_2$, may be referred
to as the total elliptic anisotropy. Here we assume that the
reaction plane is determined with high accuracy as it is achieved
by current RHIC experiments. The particle yield at RHIC can be
attributed to two main sources: the bulk matter which dominates
low $p_t$ regime and exhibits collective flow, and the (partially
suppressed) hard jets dominant at high $p_t$. Both sources
contribute to the total yield $\frac{d^2 N}{p_t d p_t d\phi}$ and
thus to the measured $v_2$.

The proposed new observable ${\mathcal C}_{FB}[p_T]$ is the
correlation of the {\em total} $V_2(p_t)$ between  forward (F) and
backward (B) rapidity bins, specifically
\begin{equation} \label{eqn_clr_def}
 {\mathcal C}_{FB}[p_T] \equiv \frac{<V_2^F\cdot V_2^B
>}{< V_2^F> \cdot < V_2^B>}
\end{equation}
where the total elliptic anisotropy in the forward, $V_2^F$, and
backward, $V_2^B$, are defined as
\begin{eqnarray} \label{eqn_clr_up}
\ave{V_2^{F/B}(p_t)} &=& \int_0^{2\pi}d\phi\cos(2\phi)
\ave{\frac{dN^{F/B}}{d\phi} },
 \\
\ave {V_2^F\cdot V_2^B \,} (p_t)
&=&\left\langle \int_0^{2\pi}d\phi\cos(2\phi)\frac{dN^{F}}{d\phi}  \right. \nonumber \\
&&\qquad \cdot \left. \int_0^{2\pi}
d\phi'\cos(2\phi')\frac{dN^{B}}{d\phi'} \right\rangle
\label{eqn_clr_dn}
\end{eqnarray}
Here we denote by $\frac{dN^{F/B}}{d\phi} \equiv
\frac{dN^{F/B}}{p_t d p_t d\phi} \cdot p_t \cdot \Delta p_t$ the
$\phi$-differential particle yield for a given $p_t$ interval in
the $F/B$ bins, respectively. The forward/backward bins $[\pm
y_{min},\pm y_{max}]$ should be chosen such that the gap
$(-y_{min},y_{min})$ is sufficiently wide to prevent a jet from
contributing to both simultaneously while still keeping both bins
within the ``flat plateau'' near mid-rapidity.

The total yield,
$\frac{dN^{F+B}}{d\phi}=\frac{dN^{F}}{d\phi}+\frac{dN^{B}}{d\phi}$,
can be decomposed into two components\cite{Gyulassy:2000gk},
the hydrodynamic flow yield and the jet-related yield, respectively:
\begin{equation}
\frac{dN^{F+B}}{d\phi} =\frac{dN^{\mathcal F}}{d\phi}{\bigg
|}_{F+B}+\frac{dN^{\mathcal J}}{d\phi}{\bigg |}_{F+B}
\end{equation}
(In the following we drop the ``$F+B$'' subscript.)

To quantify the main idea, we express the yields in the $F/B$ bins
in each event as:
\begin{eqnarray}
&&\label{eqn_yield_l}
\frac{dN^{F}}{d\phi}=(\frac{1}{2}+\eta^F)\cdot \ave{
\frac{dN^{\mathcal F}}{d\phi} }+\xi \cdot \frac{dN^{\mathcal
J}}{d\phi}
\\
&&\label{eqn_yield_r}
\frac{dN^{B}}{d\phi}=(\frac{1}{2}+\eta^B)\cdot \ave{
\frac{dN^{\mathcal F}}{d\phi} }+(1-\xi)\cdot \frac{dN^{\mathcal
J}}{d\phi}
\end{eqnarray}
In the above we have introduced three random variables to
schematically describe the fluctuations from event to event.
$\eta^{F/B}$ represent (independent) random deviations from the
average hydro flow yield in F/B bins, and satisfy
$<\eta^{F/B}>=0,<\eta^F \cdot \eta^B>=0$. The variable $\xi$
assumes values of either $0$ or $1$ in each event with equal
probability, i.e. $<\xi>=<1-\xi>=1/2$ while $\xi (1-\xi)=0$. The
physical idea is that in each event there is at most one high
$p_t$ hadron cluster contributing to the final observed
 anisotropy $V_2$ \cite{Adler:2006bw} either in the forward
or the backward rapidity bin. Calculations of dijet production at
RHIC energy show a rapid decrease of cross section with increasing
rapidity separation\cite{Szczurek:2007ep}. Also with a hard
trigger in one bin, the back jet is strongly degraded in a heavy
ion collision, inducing negligible contribution to $v_2$. In
contrast, the hydrodynamic flow (ideal or viscous) contribution to
the {\em anisotropy} is about equally split between F/B. The
so-called anti-flow\cite{Antiflow} gives rise to correlations
between rapidity $y$ and in-plane $p_x$. Its magnitude, $v_1$, at
RHIC energies is at most a few
percent\cite{Voloshin:2008dg,Antiflow} and induces a correction to
$v_2$ of the order $\delta v_2 = v_1^2<1\%$. Furthermore it
contributes equally to $v_2$ in both the F/B bins due to its
(anti-)symmetry with respect to $y=0$.

Next we introduce an interpolation function:
\begin{equation} \label{eqn_gpt}
g(p_t)= \frac{\int_0^{2\pi} d\phi \ave{\frac{d N^{\mathcal
J}}{d\phi} }  }{{\int_0^{2\pi} d\phi \ave{\frac{d N^{\mathcal
F}}{d\phi}} }+{\int_0^{2\pi} d\phi \ave{\frac{d N^{\mathcal
J}}{d\phi} } }}
\end{equation}
with $g(p_t)$ and $1-g(p_t)$ giving the relative weight of the jet
(non-collective) and hydro flow (collective) contribution to the
total, $F+B$, yield, respectively. Given this function we express
the observed $<v_2>$ in the $F+B$ bins via (\ref{eqn_vt_def}) in terms
of the flow and jet contributions:
\begin{eqnarray} \label{eqn_v2_combo}
&&<v_2(p_t)>  = \frac{\int_0^{2\pi} d\phi \cos(2\phi) {\big [}
{\big <}\frac{d N^{\mathcal F}}{d\phi}{\big >}+ {\big <}\frac{d
N^{\mathcal J}}{d\phi} {\big >} {\big ]} }{\int_0^{2\pi} d\phi
{\big [} {\big <}\frac{d N^{\mathcal F}}{d\phi}{\big >}+ {\big
<}\frac{d N^{\mathcal
J}}{d\phi}{\big >} {\big ]} } \nonumber \\
&&\qquad  = [1-g(p_t)] \cdot <v_2^{\mathcal F}> + g(p_t) \cdot
<v_2^{\mathcal J}>
\end{eqnarray}
with $<v_2^{\mathcal F/J}>$ defined as in (\ref{eqn_vt_def}) with
the corresponding hydro-flow-only/jet-only yields. Combining
(\ref{eqn_clr_up},\ref{eqn_clr_dn},\ref{eqn_yield_l},\ref{eqn_yield_r},\ref{eqn_gpt},\ref{eqn_v2_combo})
 and using the formalism of
 \cite{Koch}, we find for Eq.(\ref{eqn_clr_def}):
\begin{eqnarray}\label{eqn_clr_g}
{\mathcal C}_{FB} = \frac{(1-g)^2<v_2^{\mathcal
F}>^2+2g\left(1-g\right)<v_2^{\mathcal F}> <v_2^{\mathcal
J}>}{\left[\left(1-g\right)<v_2^{\mathcal F}>+g<v_2^{\mathcal
J>}\right]^2} \quad
\end{eqnarray}
A distinct feature is that as $g\to 0$ (hydro dominance)
${\mathcal C}_{FB}\to 1$ while ${\mathcal C}_{FB}\to 0$ in the
other limit, $g\to 1$ (jet dominance). Thus, the correlation
${\mathcal C}_{FB}[p_t]$  distinguishes between collective and
non-collective contributions and exposes the transition from the
flow to the jet regime.

In order to demonstrate the sensitivity of ${\mathcal C}_{FB}$ we
study three different scenarios within a simple model. All the
scenarios employ the same blast-wave model for the collective flow
contribution, but they differ in the $p_t$-dependence of the
jet-only elliptic anisotropy, $\ave{v_2^{\mathcal J}(p_t)}$, which
is not known for small $p_t \lesssim 3\,\rm GeV$. The relative
strength, $g(p_t)$, of flow and jet contributions is obtained by
fitting the $v_2$ data of the PHENIX collaboration
\cite{Adler:2003kt,Adler:2006bw,phenix_prel} for AuAu $0-20\%$
centrality class at $\sqrt{s}=200\, \rm GeV$ (see
Fig.\ref{fig_v2}). Scenario (a) assumes a constant
$\ave{v_2^{\mathcal J}(p_t)}$ (see Fig.\ref{fig_v2}(a)) .
Scenarios (b) (Fig.\ref{fig_v2}(b)) and (c) (Fig.\ref{fig_v2}(c))
assume a constant $\ave{v_2^{\mathcal J}(p_t)}$ for large $p_t$
which drops to half its value at $p_t=0$ for scenario (b) and to
zero for scenario (c). For both scenarios the momentum scale at
which $\ave{v_2^{\mathcal J}(p_t)}$ changes towards the values at
low $p_t$ is $p_t=2\,\rm GeV$.

The blast-wave model for the collective hydro flow contribution we
adopt from ref. \cite{Teaney:2003kp}. By parameterizing the flow
velocity field at freeze-out as in \cite{Teaney:2003kp} we have,
\begin{equation}\label{eqn_blast_wave}
{\bigg <}\frac{d N^{\mathcal F}}{d\phi}{\bigg >} \propto
\frac{1}{(2\pi)^3}\int \frac{p^\mu d\sigma_\mu}{e^{p^\mu
u_\mu/T_o}-1}
\end{equation}
The elliptic anisotropy is intrinsically built in the flow field
$u^\mu(x^{\mu})$: in $(\tau,\eta_s,r,\phi_s)$ coordinates
$u^r=\frac{r}{R}\,
u_o[1+u_2cos(2\phi_s)]\,\Theta(R_o-r)$,$u^\tau=\sqrt{1+(u^r)^2}$,
$u^{\phi_s}=u^{\eta_s}=0$. The blast-wave parameters are chosen as
$m_\pi=140\,\rm MeV$, $T_o=170\,\rm MeV$, $R_o=10\,\rm fm$,
$\tau_o=7\,\rm fm$, $u_o=0.7$, and $u_2=0.06$ which approximately
reproduces the yield at low $p_t$ \cite{Adler:2006bw}. Given this
model we calculate the flow-only $<v_2^{\mathcal F}>$ via
(\ref{eqn_vt_def}) using the yield in (\ref{eqn_blast_wave})
(black solid lines in Fig.\ref{fig_v2}(a,b,c)). We also show the
weighted flow contribution $\left(1-g\right)\cdot <v_2^{\mathcal
F}>$ as the black dash-dot lines.

In order to model the anisotropic jet attenuation we write the
jet-related yield as
\begin{eqnarray}
\ave{ \frac{d N^{\mathcal J}}{ d\phi} } \propto
\left[ 1+2 \ave{v_2^{\mathcal J}}(p_t) \cdot
\cos (2\phi)\right ]
\end{eqnarray}
where  $<v_2^{\mathcal J}>(p_t)$ parameterizes the resulting
azimuthal anisotropy. While the experimental data indicate
$<v_2^{\mathcal J}>(p_t)$ to be sizeable (7\%) and rather {\em
constant} for $p_t>6\, \rm GeV$ \cite{phenix_prel}, it is not
clear how $v_2^{\mathcal J}$ behaves at lower $p_t$. To explore
this we study the three scenarios described above which we
parameterize as follows:
\begin{eqnarray} \label{eqn_hpt_vt}
 <v_2^{\mathcal J}>=0.07 \left[ \alpha + (1-\alpha)\tanh(p_t-2\,{\rm GeV})\right]
\end{eqnarray}
The three scenarios (a), (b), and (c) correspond to $\alpha=1$,
$\alpha=\frac{3}{4}$, and $\alpha=\frac{1}{2}$, with the
respective $<v_2^{\mathcal J}>(p_t)$ plotted in Fig.\ref{fig_v2}
as colored long-dashed lines.

\begin{figure}
  \hskip 0.2in\includegraphics[width=5.5cm]{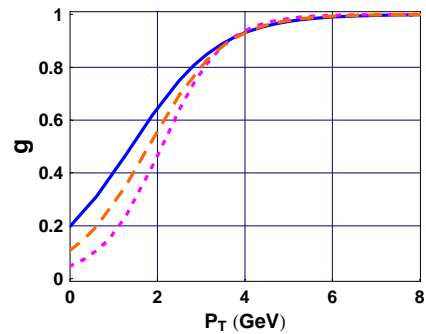}
 \caption{\label{fig_gpt}  The interpolation $g(p_t)$, Eq.(\ref{eqn_interpolation}), for scenarios (a) (solid line), (b) (long-dashed line), and (c)
(short-dashed line).}
 \end{figure}

Finally, we parameterize the interpolation (\ref{eqn_gpt}) as:
\begin{eqnarray} \label{eqn_interpolation}
g(p_t)={\big [} 1+\tanh[(p_t-P_C)/P_W] {\big ]} {\big /} 2
\end{eqnarray}
The parameters $P_C,P_W$ (in units $\rm GeV$) are obtained by
$\chi^2$ fitting of Eq.(\ref{eqn_v2_combo}) to the data for $v_2$.
The best choices for the three scenarios are: (a) $P_C=1.4$,
$P_W=2$; (b) $P_C=2.4$, $P_W=1.8$; (c) $P_C=2.4$, $P_W=1.8$. In
all cases a reasonable model description of the $v_2$ data is
established over the whole $p_t$ range, as can be seen in
Fig.\ref{fig_v2}. The resulting interpolation $g(p_t)$ are plotted
in Fig.\ref{fig_gpt}. At low to intermediate $p_t$ they differ
considerably in the magnitude of the non-collective component. Of
course the plotted curves are only examples for what may happen.
To quantify $g(p_t)$ more experimental information will be
required. We note, that an admixture of a smaller jet-$v_2$ seems
a plausible alternative to viscous corrections for a reduced $v_2$
at low to intermediate $p_t$.

\begin{figure}[tb]
    \hskip 0.1in\includegraphics[width=5.2cm]{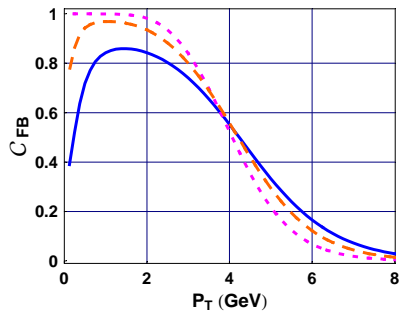}
    \vskip 0.01in
 \caption{\label{fig_clr} The correlation ${\mathcal C}_{FB}(p_t)$ for scenarios (a) (solid line),
 (b) (long-dashed line), and (c) (short-dashed line).}
\end{figure}

Given the above parametrization we can calculate the proposed
correlation ${\mathcal C}_{FB}$ using (\ref{eqn_clr_g}). The
results are shown in Fig.\ref{fig_clr}. While all three cases
produce similar $v_2$, they are readily distinguishable by
${\mathcal C}_{FB}$ in the low to intermediate $p_t$ region. The
non-monotonic structure seen for scenarios (a) and (b) is due to
interplay between a {\em non-vanishing} jet contribution and
rapidly rising hydro flow contribution to $v_2$ at low $p_t$,
which is absent in case (c). In addition, the deviation ${\mathcal
C}_{FB}$ from unity at low $p_t$ provides a measure for the
non-collective contribution to $v_2$ in this region and their
growth with increasing $p_t$. While the curves in
Fig.\ref{fig_clr} depend on the specific parametrization, they
demonstrate the sensitivity of ${\mathcal C}_{FB}$ to the
non-collectivity which may hide in $v_2$.

In summary, we have proposed to use backward-forward correlations
of the elliptic anisotropy to distinguish the collective (flow)
from non-collective (jet) contributions to $v_2$. Using a
two-component parametrization, we have studied the sensitivity of
this observable to the degree of non-collectivity. We have further
demonstrated that this observable is capable of exposing the
transition from collective flow to jets with increasing $p_t$.

Concerning an actual measurement,  one issue is that even though
the anisotropy of the hydro flow field is perfectly aligned in the
F/B bins, the actual $v_2$ of produced particles in the two bins
may deviate (both in magnitude and in orientation) from the
supposed ``$v_2$'' due to statistical fluctuation (see e.g.
\cite{Ollitrault:1993ba}). This will reduce  the correlation from
unity even for purely hydro flow, with the effect scaling as
$1/\sqrt{N_{bin}}$. The other issue relates to the low yield in
each event from intermediate to high $p_t$ which may cause the
correlation to vanish trivially. These problems may be partially
cured by selecting proper $p_t$ bin size and/or event trigger.
Also those effects are small at low $p_t$ region, where the
particle abundance is large.

To conclude, a measurement of  $\mathcal C_{FB}$ will reveal the
magnitude of the non-collective contribution to $v_2$ and the
transition pattern of $v_2$ between collective and non-collective
behavior. The centrality dependence of this transition pattern
would be of interest, as the flow and jet components scale
differently with centrality. In addition, these measurements will
help to understand the origin of the decreasing $v_2$ at
intermediate $p_t$ and constrain viscous corrections to the
hydrodynamic evolution.

We thank M. Ploskon and A. Poskanzer for discussions on the
experimental aspects of the proposed observable. This work is
supported by the Director, Office of Energy Research, Office of
High Energy and Nuclear Physics, Divisions of Nuclear Physics, of
the U.S. Department of Energy under Contract No.
DE-AC02-05CH11231.


\begin{thebibliography}{99}

\bibitem{rhic_white_paper}
  J.~Adams {\it et al.}  [STAR Collaboration],
  Nucl.\ Phys.\  A {\bf 757}, 102 (2005).
  K.~Adcox {\it et al.}  [PHENIX Collaboration],
  Nucl.\ Phys.\  A {\bf 757}, 184 (2005).
  I.~Arsene {\it et al.}  [BRAHMS Collaboration],
  Nucl.\ Phys.\  A {\bf 757}, 1 (2005).
  B.~B.~Back {\it et al.},
  Nucl.\ Phys.\  A {\bf 757}, 28 (2005).

\bibitem{Voloshin:2008dg}
  S.~A.~Voloshin, A.~M.~Poskanzer and R.~Snellings,
  arXiv:0809.2949 [nucl-ex].

\bibitem{Teaney:2000cw}
  D.~Teaney, J.~Lauret and E.~V.~Shuryak,
  Phys.\ Rev.\ Lett.\  {\bf 86}, 4783 (2001).
  P.~F.~Kolb, P.~Huovinen, U.~W.~Heinz and H.~Heiselberg,
  Phys.\ Lett.\  B {\bf 500}, 232 (2001).


\bibitem{Csernai:2006zz}
  L.~P.~Csernai, J.~I.~Kapusta and L.~D.~McLerran,
  Phys.\ Rev.\ Lett.\  {\bf 97}, 152303 (2006).

\bibitem{Lacey:2006bc}
  R.~A.~Lacey {\it et al.},
  Phys.\ Rev.\ Lett.\  {\bf 98}, 092301 (2007)

\bibitem{viscous_hydro}
  P.~Romatschke and U.~Romatschke,
  Phys.\ Rev.\ Lett.\  {\bf 99}, 172301 (2007);
  H.~Song and U.~W.~Heinz,
  Phys.\ Lett.\  B {\bf 658}, 279 (2008);
  K.~Dusling and D.~Teaney,
  Phys.\ Rev.\  C {\bf 77}, 034905 (2008).
  Z.~Xu, C.~Greiner and H.~Stocker,
  Phys.\ Rev.\ Lett.\  {\bf 101}, 082302 (2008).
H.~B.~Meyer, 
  Phys.\ Rev.\  D {\bf 76}, 101701 (2007).




\bibitem{lower_bound}
  P.~Kovtun, D.~T.~Son and A.~O.~Starinets,
  Phys.\ Rev.\ Lett.\  {\bf 94}, 111601 (2005).


\bibitem{Ollitrault:1992bk}
  J.~Y.~Ollitrault,
  Phys.\ Rev.\  D {\bf 46}, 229 (1992).



\bibitem{Shuryak:2001me}
  E.~V.~Shuryak,
  Phys.\ Rev.\  C {\bf 66}, 027902 (2002).

\bibitem{Gyulassy:2000gk}
  M.~Gyulassy, I.~Vitev and X.~Wang,
  Phys.\ Rev.\ Lett.\  {\bf 86}, 2537 (2001).
  X.~Wang,
  Phys.\ Rev.\  C {\bf 63}, 054902 (2001).


\bibitem{Liao:2008dk}
  J.~Liao and E.~Shuryak,
     Phys.\ Rev.\ Lett.\  {\bf 102}, 202302 (2009); Phys.\ Rev.\  C {\bf 75}, 054907
  (2007); Phys.\ Rev.\ Lett.\  {\bf 101}, 162302 (2008); Phys.\ Rev.\  D {\bf 73}, 014509 (2006).

\bibitem{Noncollectivity}
The Mach Cone shockwaves associated with attenuated jets, though
hydrodynamic, are relatively localized in rapidity, and thus
``non-collective'' in the present context.

\bibitem{Adler:2006bw}
  S.~S.~Adler {\it et al.} ,
  Phys.\ Rev.\  C {\bf 76}, 034904 (2007).

\bibitem{Szczurek:2007ep}
  A.~Szczurek {\it et al.} ,
  arXiv:0706.3438 [hep-ph].

\bibitem{Antiflow}
L.~P.~Csernai and D.~Rohrich, Phys. Lett. B 458 (1999) 454.
J.~Brachmann, et al. Phys. Rev. C 61 (2000) 024909.
 G.~Wang, J.
Phys. G 34 (2007) S1093. M. Lisa, et al., Phys. Lett. B 489 (2000)
287, ibid. B 496 (2000) 1.


\bibitem{Koch}
   V.~Koch,
  arXiv:0810.2520 [nucl-th].




\bibitem{Adler:2003kt}
  S.~S.~Adler {\it et al.},
  Phys.\ Rev.\ Lett.\  {\bf 91}, 182301 (2003).



\bibitem{phenix_prel}
  S.~Afanasiev {\it et al.} ,
  arXiv:0903.4886 [nucl-ex].



\bibitem{Teaney:2003kp}
  D.~Teaney,
  Phys.\ Rev.\  C {\bf 68}, 034913 (2003).

\bibitem{Ollitrault:1993ba}
  J.~Y.~Ollitrault,
  Phys.\ Rev.\  D {\bf 48}, 1132 (1993).






\end{thebibliography}
\end{document}